\documentclass[sigplan,nonacm]{acmart}
\usepackage{xcolor}
\usepackage{xspace}
\usepackage{mathtools} % upgrade from amsmath
\usepackage{bm}
\usepackage{enumitem}
\usepackage{textcomp}
\usepackage{graphicx}
\usepackage{subcaption}
\usepackage{tikz}
\usepackage[labelfont=bf]{caption}
\usepackage{bbding}
\usepackage{amsfonts}
\usepackage{tabularx}
\usepackage{multirow}
\usepackage{booktabs}
\usepackage{makecell}
\usepackage[figuresleft]{rotating}
\usepackage[flushleft]{threeparttable}
\usepackage{colortbl}

\usepackage{algorithmicx}
\usepackage{algorithm}
\usepackage[noend]{algpseudocode}

\algnewcommand\algorithmicinput{\textbf{Input:}}
\algnewcommand\Input{\item[\algorithmicinput]}
\algnewcommand\algorithmicoutput{\textbf{Output:}}
\algnewcommand\Output{\item[\algorithmicoutput]}

\usepackage[utf8]{inputenc} % Avoid insert Chinese characters
\usepackage[most]{tcolorbox} % Colored and framed text boxes
\usepackage{lipsum}
\usepackage{soul} % Colored and framed text 
\usepackage{siunitx} % For \priority command
\usepackage{pifont} % inconsolata
\usepackage{xurl} % url break line
\usepackage[noindentafter]{titlesec}
\titlespacing\section{0pt}{3pt}{3pt} %{left}{before}{after}
\titlespacing\subsection{0pt}{3pt}{3pt}
\titlespacing\subsubsection{0pt}{5pt}{5pt}

\usepackage[skip=3pt]{caption}
\usepackage{tikz}
\usepackage{amsmath}
\usepackage{makecell}
\usepackage{graphicx}
\newcommand{\sysname}{\texttt{SPPO}}

\begin{document}
%-------------------------------------------------------------------------------

%don't want date printed
\date{}

\pagestyle{plain}

\author{Qiaoling Chen$^{1,2,3}$, Shenggui Li$^{2}$, Wei Gao$^{3}$, Peng Sun$^{3,4}$, Yonggang Wen$^{2}$, Tianwei Zhang$^{2}$}

\affiliation{
  \institution{$^{1}$S-Lab, NTU \hspace{1em} $^{2}$Nanyang Technological University \hspace{1em} $^{3}$Shanghai AI Laboratory \hspace{1em} $^{4}$SenseTime}
  \country{}
}

\title{\sysname: Efficient Long-sequence LLM Training via Adaptive Sequence Pipeline Parallel Offloading}

% Practical, Accurate and Lightweight Learning-Augmented Systems with Primo

\begin{abstract} In recent years, Large Language Models (LLMs) have exhibited remarkable capabilities, driving advancements in real-world applications. However, training LLMs on increasingly long input sequences imposes significant challenges due to high GPU memory and computational demands. Existing solutions face two key limitations: (1) memory reduction techniques, such as activation recomputation and CPU offloading, compromise training efficiency; (2) distributed parallelism strategies require excessive GPU resources, limiting the scalability of input sequence length. 

To address these gaps, we propose Adaptive Sequence Pipeline Parallel Offloading (\sysname{}), a novel LLM training framework that optimizes memory and computational resource efficiency for long-sequence training. \sysname{} introduces adaptive offloading, leveraging sequence-aware offloading, and two-level activation management to reduce GPU memory consumption without degrading the training efficiency. Additionally, \sysname{} develops an adaptive pipeline scheduling approach with a heuristic solver and multiplexed sequence partitioning to improve computational resource efficiency. Experimental results demonstrate that \sysname{} achieves up to $3.38 \times$ throughput improvement over Megatron-LM and DeepSpeed, realizing efficient training of a 7B LLM with sequence lengths of up to 4M tokens on only 128 A100 GPUs.  

\end{abstract}
\maketitle

\section{Introduction}

\begin{figure}[t!]
    \centering
    \includegraphics[width=\linewidth]{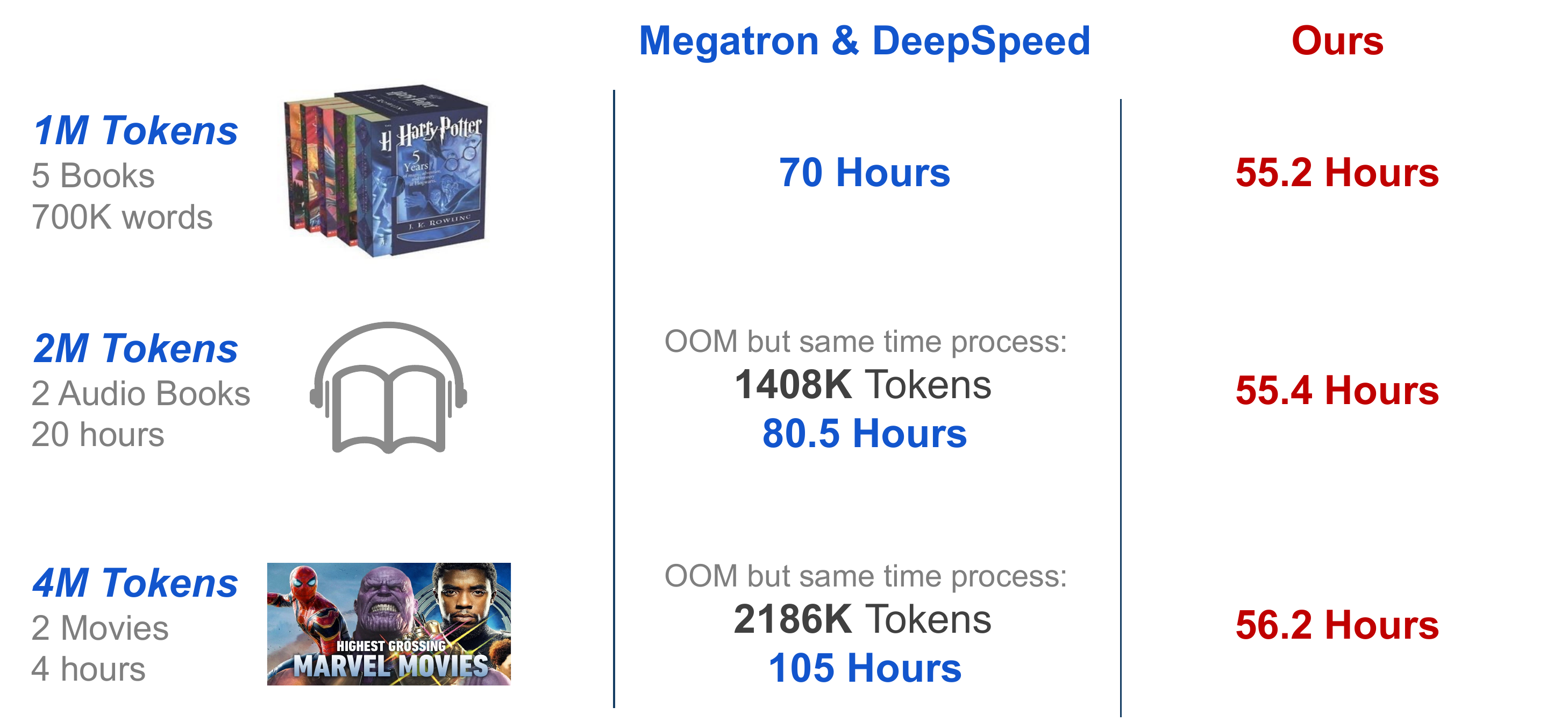}
    \caption{Performance comparisons of \sysname{} and SOTA training systems on extremely long sequences of total 1B tokens, using 32, 64, and 128 GPUs, respectively.
    } 
    \label{fig:first}
\end{figure}

Large Language Models (LLMs) have demonstrated exceptional capabilities, revolutionizing a multitude of domains including coding~\cite{codedeepseek, codellama}, image processing \cite{imageprocess, scalediffusion}, video stream analysis \cite{videosurvey,opensora}, and scientific research \cite{ai4sci1, ai4sci2}. As illustrated in Figure \ref{fig:first}, the growing length of contextual information in these applications necessitates the training of LLMs to support increasingly longer sequences, from 4K tokens \cite{alpaca, llama2} to 32K \cite{mixtral, LLaMA2-7B-32K}, 128K \cite{GPT4Report, 128Kseq}, and even millions of tokens \cite{kimi, qwen, BPT2, internevo}. Training LLMs with such long sequence lengths imposes significant demands on GPU memory and computational resources, as the activation computation and memory in training scales with the sequence length. For example, training a GPT model with the size of 7 billion and a sequence length of 4 million tokens demands approximately 16,384GB of activation memory and necessitates at least 128 NVIDIA H100 GPUs for 105 hours to process 1 billion tokens. 

% 2000 GPU hours H100 1B tokens 
% 

Therefore, many system optimizations have been proposed to improve the memory or computational resource efficiency of long-sequence LLM training. However, they are not sufficient to achieve satisfactory performance due to the following two significant gaps.

\begin{itemize}[leftmargin=*,topsep=1pt, itemsep=2pt, itemindent=8pt]
      \item {\bfseries G1}: \textsl{\bfseries Memory reduction techniques compromise the training efficiency}.
            Two prominent memory reduction techniques have been widely adopted to alleviate memory overhead in LLM training: activation recomputation \cite{sublinearmemorycost} and CPU offloading \cite{capuchin}. Activation recomputation reduces the memory footprint by recomputing the activations of certain layers during backward propagation instead of storing them. Despite the significant memory savings, the recomputing step can account for up to one-third of the gradient computation time. This overhead is particularly pronounced in long-sequence training, substantially prolonging the overall training time. CPU offloading relieves GPU memory pressure by transferring activations between GPU and CPU memory. Prior offloading techniques~\cite{TransformerEngine,ATCOffload} operate the entire sequence of activations and assume that the overhead of CPU offloading can be effectively hidden by overlapping it with GPU computation. However, as the sequence length increases, the overhead of CPU offloading grows linearly,  causing substantial delays in GPU computation and markedly degrading training efficiency. Also, existing CPU offloading techniques keep the activations of at least one layer in GPU memory, which limits their scalability for longer sequences (discussed in Section \ref{subsec:memory_reduction_techs}). Consequently, offloading the entire sequence of activations represents a \emph{coarse-grained} approach, rendering current CPU offloading methods inappropriate for long-sequence training.

      \item {\bfseries G2}: \textsl{\bfseries Distributed parallelism techniques consume excessive GPU memory and resources}.
          Many distributed parallelism strategies~\cite{loongtrain,longvila,Piper,PipeDream} have been proposed to expedite long-sequence LLM training. However, their effectiveness relies heavily on the availability of substantial GPU memory and resources. For instance, training a GPT model with 65 billion parameters and a sequence length of 4 million results in an activation memory footprint of 80TB, necessitating \textbf{over 1,024 NVIDIA H100 GPUs} to store activations and model weights. Such enormous resource requirements significantly hinder the scalability of sequence lengths in long-sequence training.

\end{itemize}

To address the substantial activation memory consumption of long sequences, prior works \cite{ringattn,DeepSpeedUlysses,BPT2,TeraPipe} have adopted \textit{sequence partitioning}, where long sequences are divided into multiple subsequences for processing. Building on them, we propose to \textit{perform CPU offloading and pipeline scheduling over the subsequences, instead of the entire sequences}. We expect such adaptions could bring two potential advantages for long-sequence LLM training: (1) Subsequence-level activation offloading benefits the overlapping of GPU computation and CPU offloading. (2) Subsequence pipeline scheduling can reduce pipeline bubbles and improve training efficiency. To our best knowledge, CPU offloading and pipeline scheduling at the subsequence granularity has not been systematically studied before. However, applying them in practice still faces challenges due to their adoption of fixed offloading policy and pipeline schedule.

\begin{itemize}[leftmargin=*,topsep=1pt, itemsep=2pt, itemindent=8pt]
            \item \textsl{\bfseries Inefficient fixed offloading.} Existing solutions commonly adopt fixed offloading policies. The length-based offloading policy \cite{nvidia3,FPDT} overlooks the computational imbalance caused by later tokens requiring more computation in the attention layers. Meanwhile, the FLOPs-based offloading policy \cite{TeraPipe} achieves computational balance but at the cost of imbalanced activation memory consumption across subsequences, causing the CPU offloading overhead to outweigh GPU computation. Furthermore, fixed subsequence offloading introduces additional tensor dependencies among subsequences, resulting in unnecessary offloading overhead.

            % In attention layers, later tokens require more computation, but the equal-length policy \cite{nvidia3,FPDT}  ignores this imbalance. The FLOPs-based policy \cite{TeraPipe} gradually shortens subsequences to achieve computational load balance among them. However, it results in \textit{imbalanced memory allocation across subsequences}, leading to inconsistent activation sizes and transmission pverh. In early subsequences, high transmission time exceeds computation, increasing GPU idle time. Additionally, finer-grained scheduling increases tensor dependencies, introducing \textit{unnecessary offloading overhead} and making fixed offloading inefficient.
      
      \item \textsl{\bfseries Inefficient fixed pipeline schedule.} The fixed pipeline schedule policy~\cite{GPipe,PipeDream,dapple,zerobuble} enforces a predetermined schedule for each subsequence, disregarding variations in memory and computational demands. More subsequences could increase kernel overhead and decrease throughput, while fewer subsequences could introduce high pipeline bubbles. Even with an optimized subsequence count \(N\), pipeline bubbles remain inevitable. For example, with \(p=4\) and \(N=16\), the bubble ratio reaches \(3/16\), highlighting inefficiencies in fixed schedules, where the total computation time includes both bubble overhead and execution time.

      % A large number of subsequences increases kernel launch overhead and decreases throughput while a small number of subsequences 
      
      % Dividing a sequence into smaller subsequences introduces a \textit{trade-off between GPU utilization and pipeline efficiency}. Reducing subsequence length increases kernel launch overhead and decreases throughput, while longer subsequences improve throughput but plateau when computation becomes bound. Additionally, subsequences also introduce \textit{inevitable pipeline bubbles}, reducing overall efficiency. Even with an optimized subsequence number \(N\), pipeline bubbles remain inevitable. For example, with \(p=4\) and \(N=16\), the bubble ratio reaches \(3/16\), highlighting inefficiencies in fixed schedules, where the total computation time includes both bubble overhead and execution time.
        
\end{itemize}

\begin{figure}[t!]
    \centering
    \includegraphics[width=\linewidth]{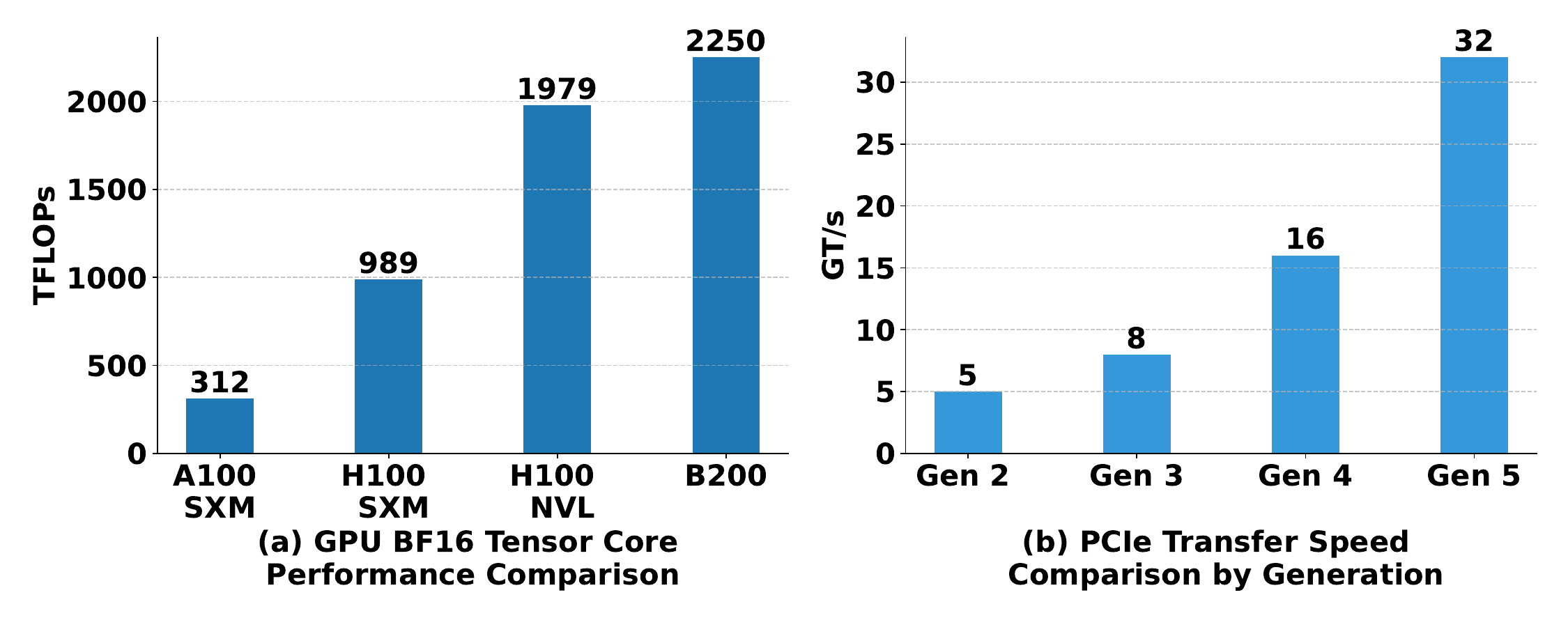}
    \caption{The evolution of GPU computing power and PCIe.} 
    \label{fig:hardware}
\end{figure}

In this paper, we propose Adaptive Sequence Pipeline Parallel Offloading (\sysname{}), a novel framework for long-sequence LLM training. It can fully exploit the potential benefits of sequence partitioning while overcoming the limitations of existing offloading policies and pipeline schedules. \sysname{} partitions long input sequences into multiple subsequences and innovatively customizes \emph{offloading} and \emph{pipeline scheduling} to optimize their memory and computational resource efficiency. First, \sysname{} designs an adaptive offloading approach to efficiently overlap the offloading of the activation of a subsequence with its computation. It consists of two key components. The first is the \textit{sequence-aware offloading} policy to mitigate imbalanced memory allocation across subsequences. This policy adaptively computes the offload ratio to maximize the overlap between the CPU offloading for the \((i-1)^{\text{th}}\) subsequence and the computation of the \(i^{\text{th}}\) subsequence. The second is the \textit{two-level activation management} strategy that retains skeletal activations (i.e., those with high access frequency) in GPU memory, thereby preserving CPU-GPU bandwidth for activations with lower access frequency.

% Moreover, different types of activations vary in access frequency patterns. Offloading frequently accessed activations can saturate the CPU-GPU bandwidth, undermining the benefits of subsequence offloading. To resolve this, we propose a \textit{two-level activation management} strategy that retains skeletal activations (i.e., those with high access frequency) in GPU memory, thereby preserving CPU-GPU bandwidth for activations with lower access frequency.

Second, we design an adaptive pipeline schedule to pipeline the computation of one subsequence in a given layer with the computation of the previous subsequence in the subsequent layer. To strike an optimal balance between GPU utilization and pipeline efficiency, we develop a \textit{heuristic solver} to determine the ideal number of subsequences. Combined with the offloading ratio in adaptive offloading, this enables us to further improve both memory efficiency and training efficiency for long-sequence LLM training. Nevertheless, the \textit{heuristic solver} is not always capable of eliminating resource bubbles in certain scenarios. To handle it, we introduce a \textit{multiplexing sequence partitioning} to deliver a fine-grained partition of adjacent subsequences, further reducing the resource bubbles without sacrificing memory efficiency.

Our contributions are summarized as follows:
\begin{itemize}[leftmargin=*,topsep=1pt, itemsep=2pt, itemindent=8pt]
    \item We propose and implement the \sysname{} framework to partition long sequences into multiple subsequences. By tailoring dedicated offloading techniques and optimizing the pipeline schedule, \sysname{} significantly improves both memory and computational efficiency for long-sequence LLM training. 
    \item We introduce an adaptive offloading approach, which comprises of sequence-aware offloading and two-level activation management, to maximize the overlap between CPU offloading and GPU computation.
    \item We design an adaptive pipeline schedule that incorporates a heuristic solver and multiplexed sequence partitioning, enhancing training efficiency without compromising the benefits from adaptive offloading. 
    \item We evaluate \sysname{} through extensive experiments, demonstrating up to a \(3.38 \times\) throughput improvement over Megatron-LM and DeepSpeed. Moreover, \sysname{} enables efficient training of a 7B LLM with sequence lengths of 1M, 2M, and 4M on only 32, 64, and 128 NVIDIA Ampere GPUs.
\end{itemize}

\begin{figure*}[t!]
    \centering
    \includegraphics[width=\linewidth]{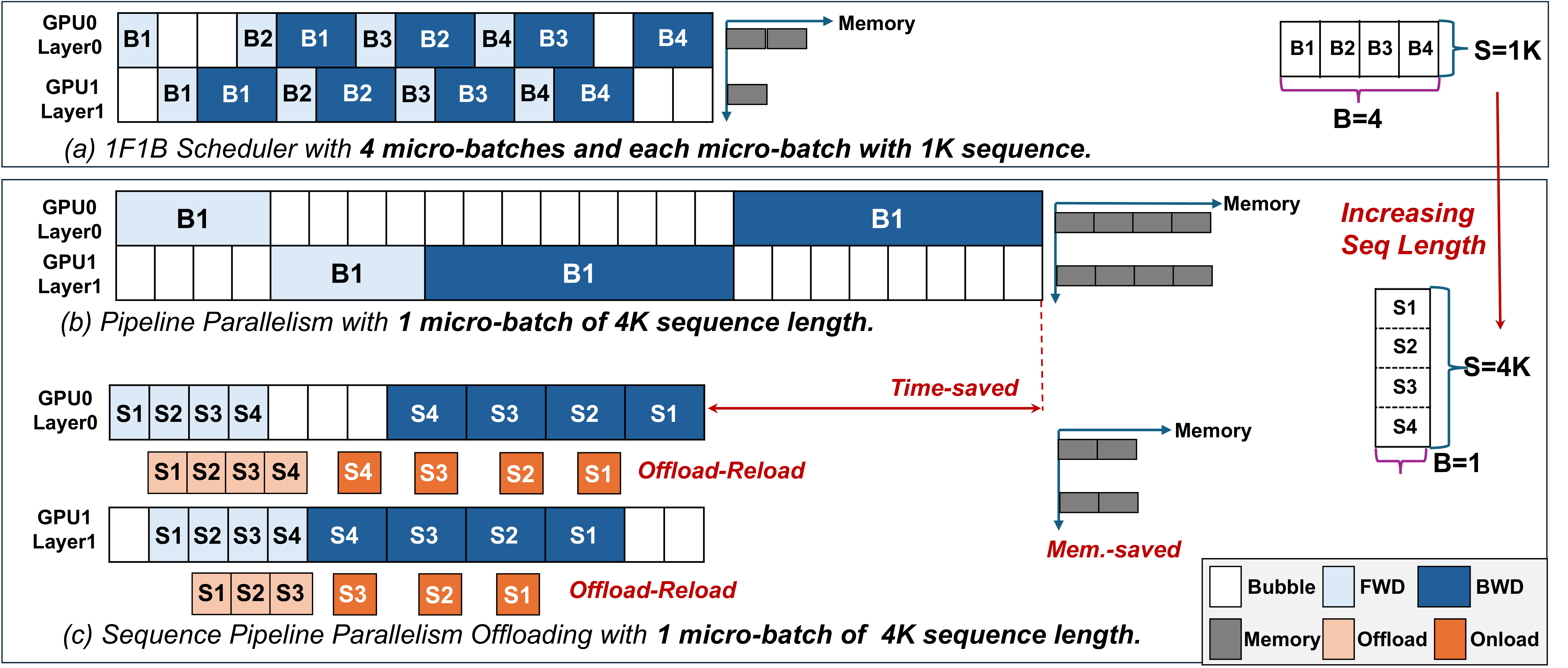}
    \caption{Illustration of the pipeline parallelism scheduling with the increasing sequence length and our sequence pipeline parallel offloading scheduling.} 
    \label{fig_pipelilnes}
\end{figure*}

\begin{table}[h]
    \centering
    
    \begin{tabular}{@{}l@{\hspace{0.15cm}}l@{\hspace{0.15cm}}|@{\hspace{0.15cm}}l@{\hspace{0.15cm}}l@{}}
   \toprule
        $B$ & batch size &   $PP$ & pipeline parallel size  \\
        $H$ & hidden dimension & $SP$ & sequence parallel size  \\
        $S$ & sequence length& $N$ & \#subsequences \\
        $M_m$ & memory of model & $s_i$ & subsequence $i$ \\
        $M_a$ & memory of activation & $\alpha$ & offloading ratio \\
        $BW$ & bandwidth & $h$ & hidden state \\
        \hline
    \end{tabular}
\caption{Notations used in this paper.}
    \label{tab:notation}
\end{table}

\section{Background}
Despite the substantial memory optimizations in attention layers achieved by FlashAttention \cite{flashatten1,flashatten2} and linear attention \cite{mamba1,Mamba2}, the linear scaling of activation memory during long-sequence LLM training still results in prohibitively high GPU memory usage. To address such overhead while maintaining computational efficiency, two primary techniques have emerged: memory reduction and distributed parallelism. However, they struggle to maintain a balance between memory efficiency and computational resource efficiency simultaneously. Table \ref{tab:notation} illustrates the notations used in this paper.

\subsection{Memory Reduction Techniques}
\label{subsec:memory_reduction_techs}
Limited GPU memory is a critical bottleneck of training LLMs with long-sequences. To address this, the LLM community has adopted activation recomputation and offloading to reduce GPU memory consumption caused by activations.

\noindent\textbf{Activation recomputation} \cite{tianqirecomp,DTR,nvidia3}, also referred to as activation checkpointing, reduces GPU memory usage by selectively discarding the activations of certain layers instead of retaining all intermediate activations. During backpropagation, the missing activations are recomputed on the fly. While activation recomputation alleviates the GPU memory consumption, the recomputation process brings additional 
computational overhead, degrading the overall training efficiency. For example, when we employ activation recomputation for all layers during long-sequence training, the cost of recomputing activations in backpropagation adds an extra 1/3 to the total computational time.

\noindent\textbf{CPU Offloading} \cite{TransformerEngine,vdnn,ATCOffload} relieves the GPU memory burden by transferring activations from GPU memory to CPU and loading back to the GPU memory on demand. Unlike activation recomputation, the ffloading process does not involve GPU computation. Existing offloading techniques~\cite{zeroOffload,TransformerEngine,ATCOffload} emphasize the ineligible cost of transferring activations between the GPU and CPU, and they attempt to maximize the overlap between the transferring process and GPU computation. However, these methods typically focus on offloading the activations of the entire input sequence for one or more layers to the CPU. During long-sequence training, the GPU memory consumption of activations grows linearly with the sequence length, quickly saturating the CPU-GPU bandwidth. As a result, the offloading process fails to effectively overlap with GPU computation, leading to training inefficiency. %Hence, existing CPU offloading techniques struggle to preserve training efficiency through overlapping. 
Worse still, these offloading techniques need to retain the activations of the entire input sequence for at least one layer in GPU memory, which significantly limits the scalability of the sequence length. For example, when training a GPT-7B LLM with a GPU equipped with 80 GB memory, retaining activations for just one layer results in a maximum sequence length of only 128K tokens.

\subsection{Distributed Parallelism Techniques}
\label{subsec:distributed_parallel}
Distributed parallelism is widely adopted to expedite LLM training. It includes several key techniques, including data parallelism, tensor parallelism, pipeline parallelism, and sequence parallelism. Hybrid parallelism strategically combines multiple techniques to further accelerate training.

\noindent \textbf{Data Parallelism (DP)} \cite{PyTorchDistributed,AMSP} segments the input data into smaller shards and distributes these shards across multiple GPUs along the batch dimension. Each GPU independently performs gradient computation, followed by gradient synchronization across GPUs with an \texttt{all-reduce} operation.

\noindent \textbf{Tensor Parallelism (TP)} \cite{Megatron-LM} involves splitting model parameters of certain layers along certain dimensions across GPUs to parallelize the training process.  In Megatron-LM \cite{Megatron-LM}, TP is employed to split linear layers by row or column dimensions, effectively reducing the GPU memory footprint of model weights while enhancing training efficiency.

\noindent \textbf{Sequence Parallelism (SP)} 
\cite{nvidia3,BPT2,ringattn,DeepSpeedUlysses,loongtrain} divides the sequence along the sequence dimension and distributes subsequences across GPUs for parallel computation. Megatron-SP \cite{nvidia3} leverages \texttt{all-gather} and \texttt{reduce-scatter} collectives to amortize the attention computations among GPUs and aggregate the computation results, thereby increasing the attention computation speed of long-sequence training.

\noindent \textbf{Pipeline Parallelism (PP)} \cite{PipeDream,GPipe} partitions an LLM into several stages, distributing these pipeline stages across GPUs. During training, consecutive stages need to exchange gradients and activations. However, this dependency can result in significant GPU idle time, where computation waits for communication to complete, a phenomenon known as \textit{pipeline bubbles}. Early works like GPipe \cite{GPipe} reduce pipeline bubbles via increasing the number of concurrent microbatches, albeit at the expense of higher peak memory usage. Subsequent works, including 1F1B \cite{PipeDream} and TeraPipe \cite{TeraPipe}, realize bubble reduction through careful scheduling policies.

\begin{table}[]
\begin{tabular}{@{}ccccccc@{}}
\hline
Model & L & h & a & $M_{m}$ & $M_{a}$ & \#GPUs \\
\hline
GPT-7B & 32 & 4096 & 32 & 120 & 16384 & 512+ \\
GPT-13B & 40 & 5120 & 40 & 234 & 52600 & 660+ \\
GPT-65B & 80 & 8192 & 64 & 1200 & 81920 & 1024+ \\
\hline
\end{tabular}
\caption{Training LLMs of varied sizes with sequence length of 4M tokens: memory footprint of model and activation in GB, and minimal number of A100 (80GB) GPUs required.}
\label{tab:min_num_gpus}
\small
\end{table}

These techniques can accelerate LLM training but require excessive GPU resources. As shown in Table~\ref{tab:min_num_gpus}, training sequences with lengths of 4 million tokens can require hundreds or even thousands of GPUs. Such high GPU demand primarily arises from the massive memory required for activations. For instance, training a GPT-65B model with an input sequence length of one million tokens under sequence parallelism consumes 28.8 TB of activation memory, requiring at least 288 A100 GPUs to accommodate it. This underscores the critical need for an efficient approach to reduce memory consumption while maintaining training efficiency.

\section{Motivation and Challenges}
We first demonstrate that subsequence offloading and pipeline scheduling can provide potential memory and computational advantages over processing the entire sequence in long-sequence LLM training. Next, we investigate the impact of current subsequence offloading and pipeline schedule policies on the training efficiency. Unfortunately, neither of them can unlock the benefits offered by subsequence partitioning.

\begin{figure}[t!]
    \centering
    \includegraphics[width=\linewidth]{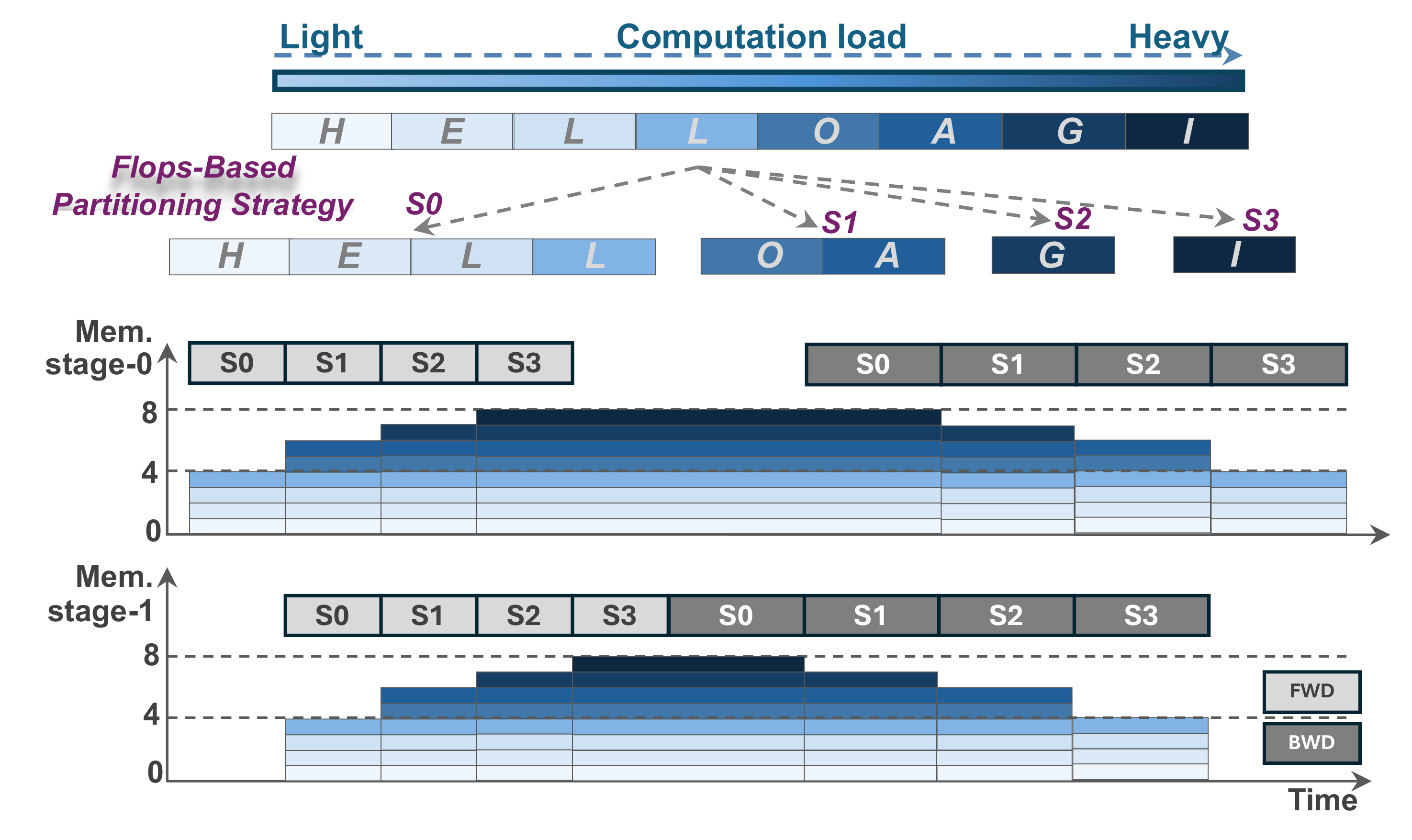}
    \caption{\textbf{Background \& motivation.} Imbalanced computation across subsequences with its FLOPs-based offloading policy and the following memory allocation in one step.} 
    \label{fig_comp_balanced}
\end{figure} 

\subsection{Potential Benefits}
\label{subsec:benefit-of-subsequence}
A promising approach to optimizing long-sequence LLM training is to split a long sequence into many subsequences \cite{ringattn,DeepSpeedUlysses,BPT2,TeraPipe}. Thus, we can opportunistically employ CPU offloading and pipeline scheduling over \emph{subsequences} to unleash the potential memory and computation advantages.

% The primary performance bottleneck of long-sequence LLM training stems from the existence of substantial activation memory consumption caused by the long input sequence. Thus, we attempt to split a long input sequence into many subsequences and employ CPU offloading and pipeline scheduling over subsequences to explore the potential memory and computation advantages. 

\begin{itemize}[leftmargin=*,topsep=1pt, itemsep=2pt, itemindent=8pt]
      \item {\bfseries Benefit 1}: \textsl{\bfseries Subsequence offloading}. Subsequence partition can optimize the overlapping of GPU computation and CPU offloading in two aspects. First, by partitioning the sequence into subsequences, the costs of CPU offloading and GPU computation become dependent on the length of the subsequences rather than the entire sequence.
      
      % As discussed in Section~\ref{subsec:memory_reduction_techs}, the overhead of offloading activations for the entire input sequence cannot overlap well with the GPU computation. With 
      
      % Despite with the difficulty of hid the overhead of offloading activations for the entire input sequence with the GPU computation, we can tune the subsequence length to identify the optimal overlap between GPU computation and CPU offloading.

      % As discussed in Section~\ref{subsec:memory_reduction_techs}, it is quite challenging to fully hide the overhead of offloading the activations for entire input sequences. Subsequence partition can achieve concurrent overlapping of GPU computation and CPU offloading at the granularity of subsequences. This partitioning enables the hiding of offloading overhead within GPU computation for two aspects. First, by partitioning the sequence into subsequences, GPU memory consumption becomes dependent on the length of the subsequences rather than the entire sequence. By flexibly tuning the subsequence length, we can identify the optimal overlap between GPU computation and CPU offloading.
       
\end{itemize}

Second, the CPU offloading overhead can be further reduced with the growing CPU-GPU bandwidth. As shown in Figure~\ref{fig:hardware},  the growth rate of CPU-GPU bandwidth has outpaced computational gains by a factor of over 30. This significant improvement in bandwidth makes it promising to hide the overhead of offloading subsequence activations to the CPU without compromising training efficiency.

% We observe that the growth of CPU-GPU bandwidth outpaces the improvement in GPU computational capability. For instance, the half-precision performance of Nvidia H100 and B200 is 1979 and 2240 TFLOPs, respectively, representing only a 13\% increase in computational capability. In contrast, the transition from PCIe 3.0 to PCIe 5.0 has provided a 4 \(\times\) improvement in CPU-GPU bandwidth. Increased CPU-GPU bandwidth makes it promising to hide the overhead of offloading subsequence activations to the CPU without compromising training efficiency.

% the growth of CPU-GPU bandwidth has significantly outpaced the improvement in GPU computational capability. For instance, the half-precision performance of Nvidia H100 and B200 is 1979 and 2240 TFLOPs, respectively, representing only a 13\% increase in computational capability. In contrast, the transition from PCIe 3.0 to PCIe 5.0 has provided a 4 \(\times\) improvement in CPU-GPU bandwidth. This bandwidth enhancement surpasses computational gains by a factor of over 30. Leveraging this, increased CPU-GPU bandwidth makes it promising to hide the overhead of offloading subsequence activations to the CPU without compromising training efficiency.

\begin{itemize}[leftmargin=*,topsep=1pt, itemsep=2pt, itemindent=8pt]
      \item {\bfseries Benefit 2}: \textsl{\bfseries Subsequence 
      pipeline schedule}. Incorporating subsequence partitioning with pipeline parallelism can reduce pipeline bubbles and improve training efficiency. We use a 1F1B pipeline scheduler~\cite{PipeDream} to demonstrate this in Figure~\ref{fig_pipelilnes}. The 1F1B scheduler decomposes input batches into multiple micro-batches and alternates forward/backward passes as illustrated in Figure~\ref{fig_pipelilnes} (a). However, under long-sequence inputs, the micro-batch count collapses to one, thus 1F1B degenerating to the naive pipeline schedule shown in Figure~\ref{fig_pipelilnes} (b), thereby incurring larger pipeline bubbles.  As illustrated in Figure~\ref{fig_pipelilnes} (c), subsequence partitioning enables the pipelining of GPU computations for one subsequence with another. This approach reduces pipeline bubbles and maximizes resource utilization, offering a more efficient pipeline schedule for long-sequence training.

      % subsequence partitioning facilitates the parallelization of computations for one subsequence on a given layer with the previous subsequence on the subsequent layer of the model. We consider a pipeline schedule example with an input sequence of $T=4\text{K}$ tokens. For a transformer layer, the input tensor maintains dimensions $(B, S, H)$, where $B$ denotes the batch size, $S$ is the sequence length, and $H$ is the hidden state dimension. The 1F1B scheduler decomposes input batches into $n=4$ micro-batches with $S=1\text{K}$ and alternates forward/backward passes as illustrated in Figure~\ref{fig_pipelilnes} (a). However, under long-sequence inputs ($S=4\text{K}$), the micro-batch count collapses to $n=1$, thus 1F1B degenerating to the naive pipeline schedule shown in Figure~\ref{fig_pipelilnes} (b), thereby incurring larger pipeline bubbles.  As illustrated in Figure~\ref{fig_pipelilnes} (c), subsequence partitioning enables the pipelining of GPU computations for one subsequence with another. This approach reduces pipeline bubbles and maximizes resource utilization, offering a more efficient pipeline schedule for long-sequence training.

      % Subsequence partitioning is naturally compatible with pipeline parallelism to enhance the efficiency of long-sequence training. First, pipeline parallelism is an effective technique that reduces GPU memory consumption by distributing model parameters across devices. It is particularly well-suited for long-sequence training to optimize its memory usage. 
\end{itemize}

% Second, subsequence partitioning facilitates the parallelization of computations for one subsequence on a given layer with the previous subsequence on the subsequent layer of the model. We consider a pipeline schedule example with an input sequence of $T=4\text{K}$ tokens. For a transformer layer, the input tensor maintains dimensions $(B, S, H)$, where $B$ denotes the batch size, $S$ is the sequence length, and $H$ is the hidden state dimension. The 1F1B scheduler decomposes input batches into $n=4$ micro-batches with $S=1\text{K}$ and alternates forward/backward passes as illustrated in Figure~\ref{fig_pipelilnes} (a). However, under long-sequence inputs ($S=4\text{K}$), the micro-batch count collapses to $n=1$, thus 1F1B degenerating to the naive pipeline schedule shown in Figure~\ref{fig_pipelilnes} (b), thereby incurring larger pipeline bubbles.  As illustrated in Figure~\ref{fig_pipelilnes} (c), subsequence partitioning enables the pipelining of GPU computations for one subsequence with another. This approach reduces pipeline bubbles and maximizes resource utilization, offering a more efficient pipeline schedule for long-sequence training.

\begin{figure}[t!]
    \centering
    \includegraphics[width=\linewidth]{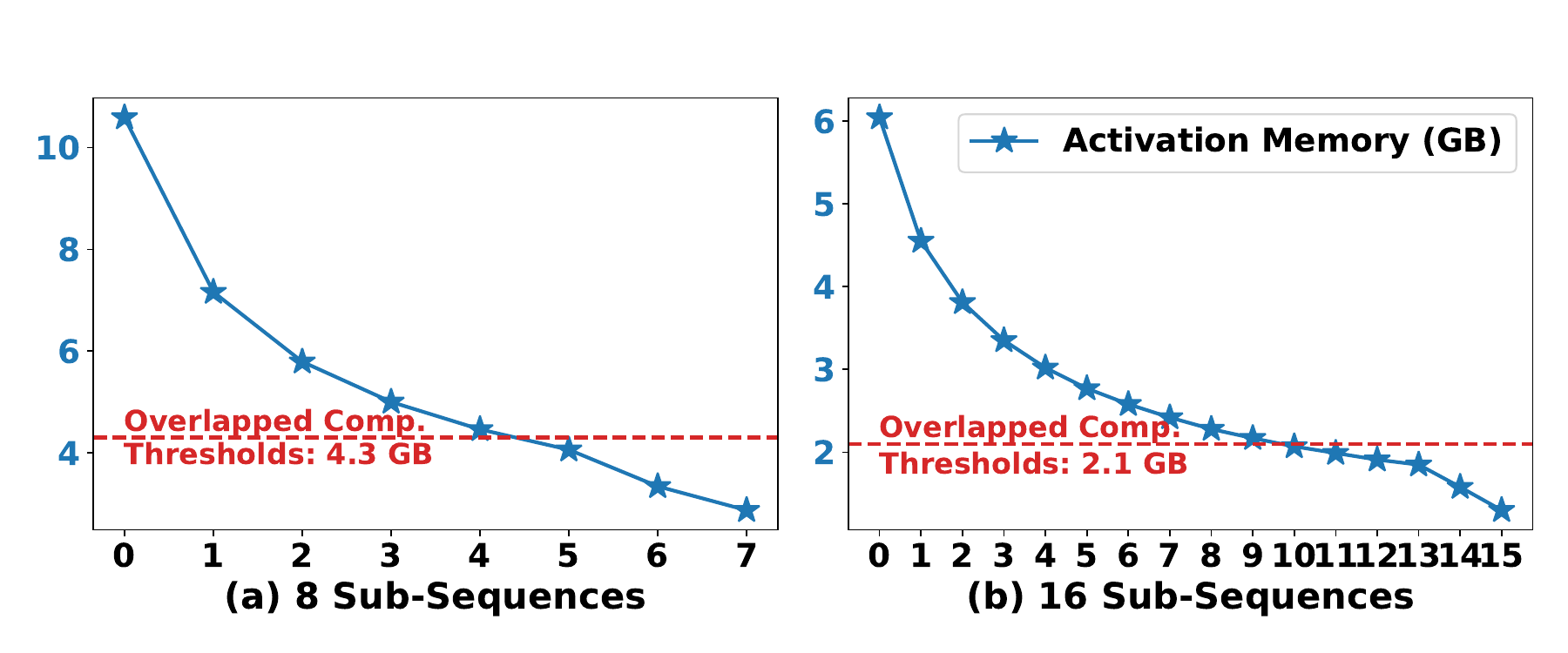}
    \caption{Activation memory allocation across subsequences when applying the optimal strategy of partitioning the sequence to 8 and 16 subsequences, respectively. The model is LLaMA-65B, and the sequence length is 128K. } 
    \label{fig_imbalaced_offload}
\end{figure} 

\begin{figure}[t!]
    \centering
    \includegraphics[width=\linewidth]{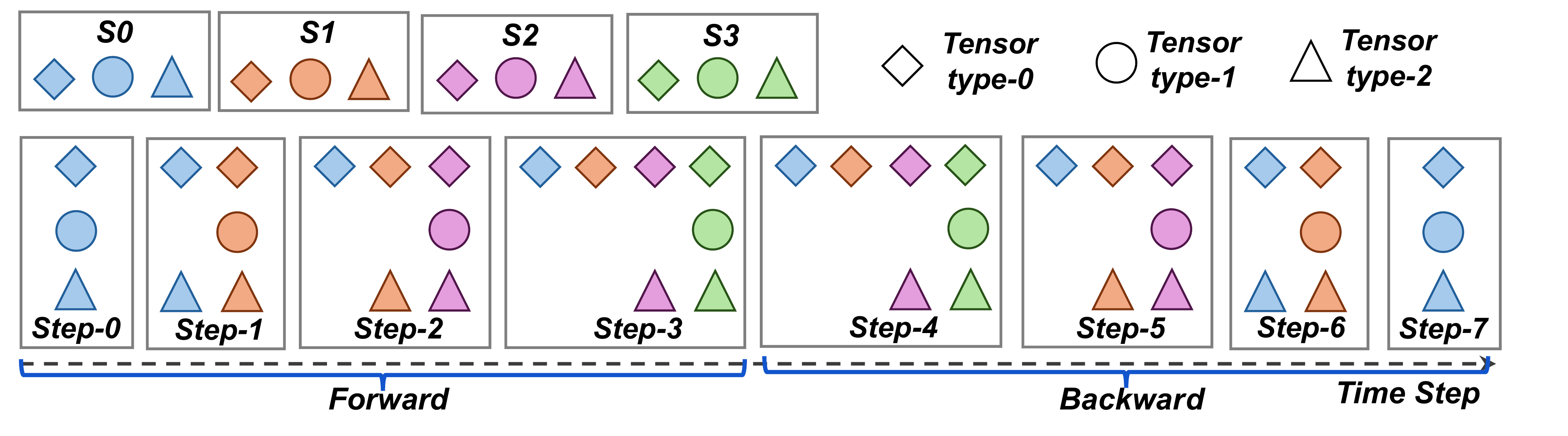}
    \caption{ Transformer-based model tensor access timeline.} 
    \label{fig_unnecessary_offload}
\end{figure}

\subsection{Challenge 1: Inefficient Fixed Offloading} 
\label{subsec_offloading_inefficient}
Following previous works, a common practice is to adopt a fixed offloading policy, which can be implemented in two ways: (1) in a length-based policy \cite{ringattn,BPT2,FPDT,DeepSpeedUlysses,nvidia3}, a consistent length is fixed across partitioned subsequences; (2) in a FLOPs-based policy \cite{TeraPipe,seq1f1b}, the computational FLOPs are kept consistent among the partitioned subsequences. However, we emphasize that these fixed offloading strategies do not take into account the efficiency of the trade-offs between offloading and computation, as well as the access frequency patterns of the activation tensor, resulting in suboptimal overlapping between CPU offloading and GPU computation.

% some efficiency problems.

\noindent\textbf{Imbalanced GPU computation across subsequences.} For the attention layer, the computational load for later token positions in a sequence is heavier than that for earlier tokens \cite{distflashattn}. However, dividing the input sequence into multiple equal-length subsequences \cite{FPDT,nvidia3} overlooks this computational load imbalance. A more efficient partitioning strategy would involve starting with longer subsequences and gradually transitioning to shorter ones. 

\noindent\textbf{Imbalanced memory allocation across subsequences.} The FLOPs-based offloading policy \cite{TeraPipe} ensures a consistent computational load across subsequences, as illustrated in Figure \ref{fig_comp_balanced}. However, due to the inherent computation imbalance in the attention layer, subsequence lengths vary under this policy. Since the transmission overhead caused by offloading correlates linearly with the subsequence length, the FLOPs-based offloading policy cannot overlap well between the CPU offloading and GPU computation, leading to increased GPU idle time and reduced training efficiency. As shown in Figure \ref{fig_imbalaced_offload}, while the computational cost remains balanced under the FLOPs-based offloading policy, the size of the activations using a fixed offloading strategy varies significantly. For 8-subsequence and 16-subsequence strategies, the activation size ranges from 10.59 GB to 2.87 GB and from 6.04 GB to 1.3 GB, respectively. Correspondingly, the transmission time ranges from 821 ms to 254 ms and from 435 ms to 90 ms, respectively. Notably, for early subsequences, the transmission time exceeds the computation time, resulting in substantial offloading overhead. This highlights the need for a more fine-grained offloading policy that adapts the sequence length to maximize the overlap between computational load and activation offloading.

\noindent\textbf{Unnecessary offloading overhead.} 
Previous works have proposed to make memory offloading decisions based on the completion time of the forward pass, either at the granularity of individual layers \cite{layerwiseoffloading, FPDT, Memo} or batches \cite{ATCOffload}. These approaches work because the computation graph is traversed either layer-by-layer or batch-by-batch, and the intermediate activation tensors within a layer or batch are independent of those in later layers or batches. However, if we instead schedule computation at a finer granularity, such as subsequence, the tensor dependencies between subsequences increase, making offloading more complex, and the fixed offloading strategy is inefficient. To better understand the access patterns across subsequences, we profile three types of tensors at one iteration, tracking their access timestamps during both the forward and backward passes. As shown in Figure \ref{fig_unnecessary_offload}, Type-0 tensors (from $s_0$) are accessed continuously, while Type-1 tensors appear once during the forward pass and once during the backward pass. By exploiting these regular access patterns, we can derive a more intelligent and fine-grained memory offloading strategy by targeting specific tensors for offloading. For example, offloading Type-1 tensors is more efficient than offloading Type-0 and Type-2 tensors, which would otherwise incur high swapping overhead when offloaded at the subsequence granularity.

\subsection{Challenge 2: Inefficient Fixed Pipeline Schedule} 
\label{subsec_comp_inefficient}

The fixed pipeline schedule policy, widely adopted in existing works \cite{GPipe,PipeDream,dapple,zerobuble}, follows a predetermined schedule for each subsequence without taking into account the varying memory and computational demands across subsequences.

\begin{figure}[t!]
    \centering
    \includegraphics[width=\linewidth]{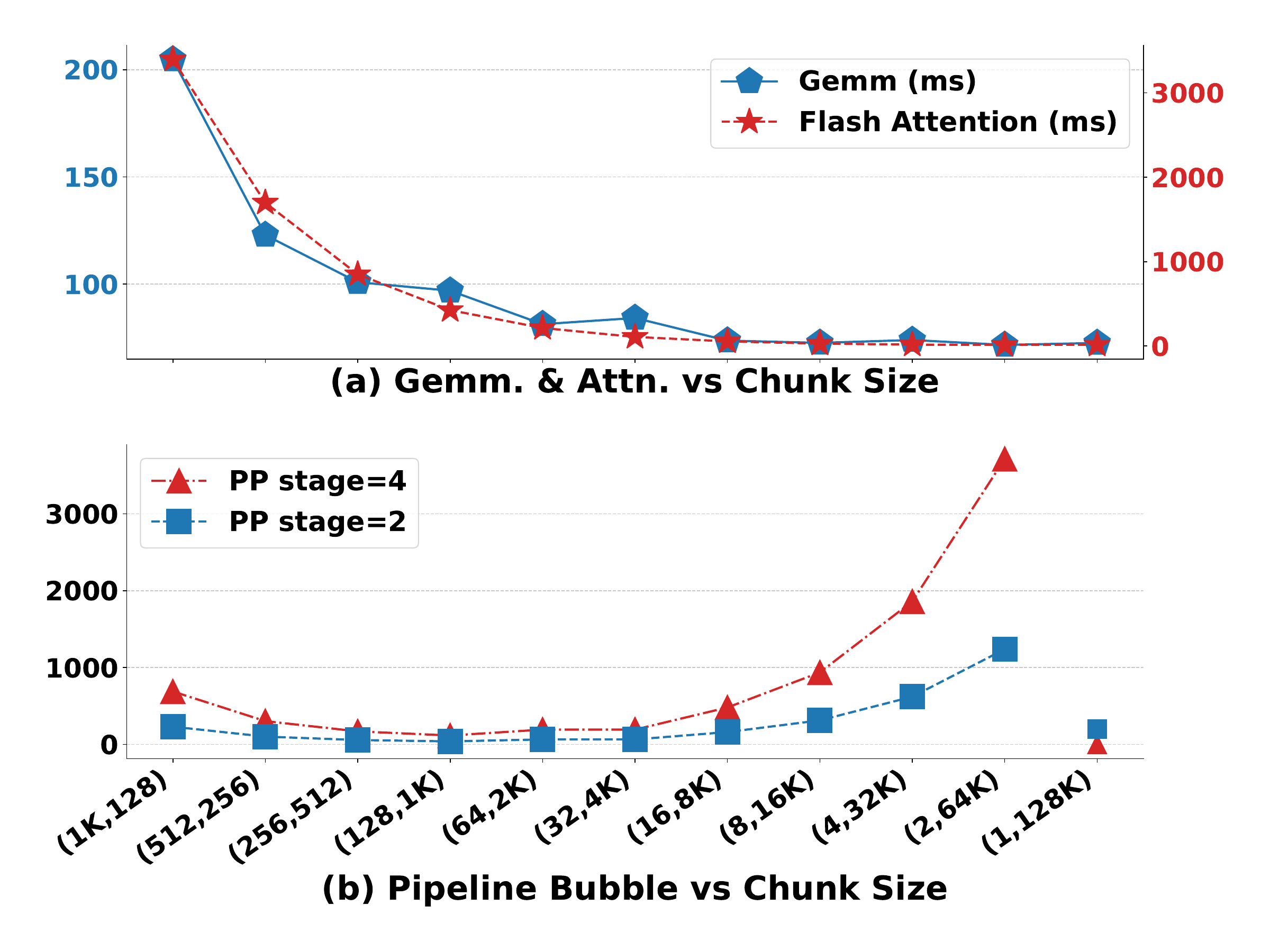}
    \caption{The relationship between subsequence length and overall forward propagation and bubble time for one transformer layer with a hidden dimension of 4096 and a fixed sequence length of 128K. The x-axis $(N,s)$ represents a sequence of 128K length that is partitioned into $N$ parts, and each part has $s$ tokens.} 
    \label{fig_trade_off_chunksize}
\end{figure}

\noindent\textbf{Trade-off on subsequence length.}
Splitting a sequence into smaller subsequences introduces a trade-off between GPU utilization and pipeline efficiency. 
As demonstrated in Figure~\ref{fig_trade_off_chunksize} (a), considering a single layer with a hidden dimension of 4096 and a constant sequence length of 128K, the overall forward propagation time can be mathematically represented as the product of the number of subsequences and the time per subsequence. As the subsequence length is reduced, the total runtime escalates due to an increase in the number of subsequences and associated kernel launch overheads. Moreover, shorter subsequences lead to reduced throughput as each pipeline stage processes only a finite amount of data, thus underutilizing GPU resources. While longer subsequences are essential for higher throughput, there exists a threshold beyond which, augmenting the subsequence length ceases to enhance overall performance. This occurs as computation operators, such as attention mechanisms and generalized matrix multiplications (GEMM), stay in a compute-bound state, thereby stabilizing the total latency. On the other hand, while longer subsequences improve throughput, they introduce more pipeline bubbles, ultimately reducing overall training efficiency, as shown in Figure \ref{fig_trade_off_chunksize} (b).  
Hence, the key is to determine an \emph{optimal} number of subsequences \(N\) to maximize the training efficiency.
\noindent\textbf{Inevitable bubble overhead.} Although it is possible to find an optimal balance between GPU utilization and pipeline bubbles, efficiency inevitably suffers from the presence of pipeline bubbles. To quantify this effect in this degradation pipeline parallelism schedule, we define \(p\) as the number of pipeline stages and \(F(N)\) as the time to execute the forward and backward passes for all \(N\) subsequences in a sequence. The time for a single subsequence is thus \({F(N)}/{N}\). The total pipeline bubble time \(t_b\) and bubble ratio \(R_b\) are computed as:
\[
t_{b} = (p - 1)\cdot \frac{F(N)}{N}, R_b = \frac{p - 1}{N}.
\]
Thus, the total computation time is:
\[
T = t_{b} + F(N) = \frac{(p - 1 + N)}{N} \cdot F(N).
\]

To minimize the overall computation time \(T\), we can determine an optimal value for \(N\). However, the pipeline bubble is inevitable by the fixed $N$. For example, when \(p=4\) and \(N=16\), the bubble ratio reaches a maximum of \(3/16\).

\begin{figure*}[t!]
    \centering
    \includegraphics[width=\linewidth]{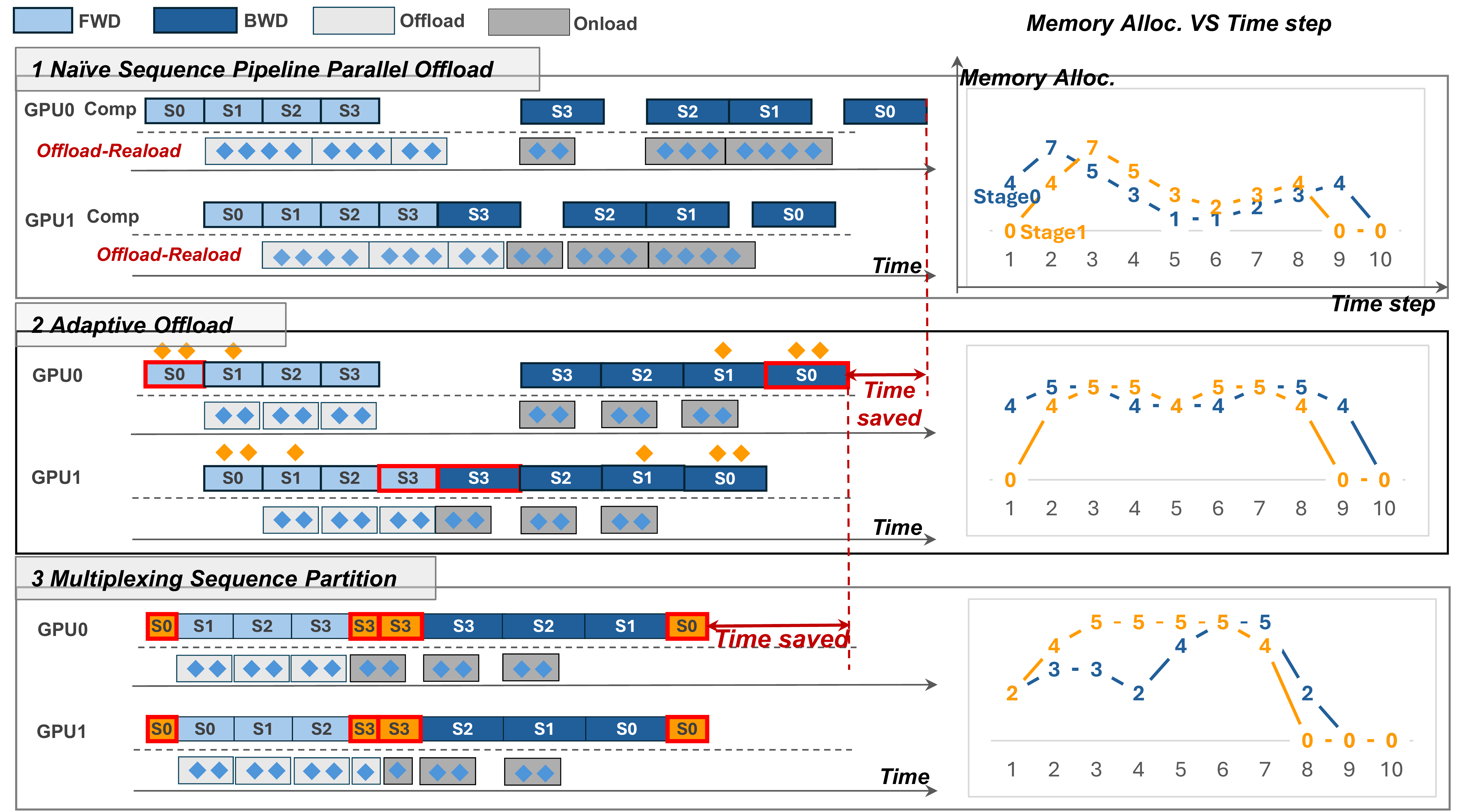}
    \caption{System Overview. The top part illustrates a minimal case of \sysname{}, where all subsequence activations are fully offloaded to host memory. The middle part demonstrates an adaptive offloading strategy that optimizes the minimal case, incorporating sequence-aware offloading and a two-level activation management scheme. The bottom further enhances efficiency through multiplexed sequence partitioning, achieving an almost bubble-free pipeline. The right plot visualizes the GPU activation memory allocation over time steps. } 
    \label{fig_system_overview}
\end{figure*}

% To keep the bubble fraction small, we need $n \gg p$; however, an overly large $n$ implies many small chunks and thereby poor GPU utilization.

% Going deeper into the intricacies of TeraPipe, the efficiency of its implementation relies heavily on the amount of device idle time referred to as pipeline bubbles. Due to the dependency between layers, bubbles seem inevitable. A prominent early work to address this issue is GPipe, which attempts to reduce the bubble ratio by increasing the number of concurrent batches in the pipeline. Motivated by this, we can ... However,,,,,

\section{System Overview}
To enhance the memory and computational performance for long-sequence training, we propose Sequence Pipeline Parallel Offload (\sysname{}), a novel LLM framework that introduces innovative sequence partitioning, offloading, and pipeline scheduling customization. As mentioned in Sections \ref{subsec_offloading_inefficient} and \ref{subsec_comp_inefficient}, existing offloading and pipeline solutions exhibit several challenges in processing subsequences. Therefore, in \sysname, we introduce two approaches to address them, respectively.

%Specifically, we split an input sequence $x_{1}, \dots, x_{L}$  into $s_{1}, \dots, s_{N}$, where each subsequence $s_{i}$  consists of tokens $\bigl(x_{l}, x_{l+1}, \dots, x_{r}\bigr)$, parallelizing the computation of the current subsequence in the current layer with the previous subsequence in the next layer of the model. We introduce how \sysname{} performs sequence partitioning and offloading. First, we apply the FLOPs-based partitioning policy for balanced computation across pipeline stages instead of the length-based partition policy (Section \ref{subsec_offloading_inefficient}). Second, our offloading policy should follow two principles: (1) Reloading starts at the beginning of the previous subsequence backward. With this scheduling, we can overlap the swap transmission time of the $(i-1)^{th}$ subsequence with the current $i^{th}$ computation, which can reduce the end-to-end training time while also significantly reducing memory consumption per pipeline stage. Figure \ref{fig_pipelilnes} (d) illustrates the simple paradigm of \sysname{}. However, our adopted sequence partitioning and offloading still pose two significant challenges to fully capitalize on the benefits of sequence partitioning. 

\begin{itemize}[leftmargin=*,topsep=1pt, itemsep=2pt, itemindent=8pt]
      \item \textsl{\bfseries Adaptive Offloading}. The top part of Figure \ref{fig_system_overview} shows the inefficient fixed offloading (Challenge 1), which has two pipeline stages and four subsequences and employs a full offloading strategy. The offloading time of $s_0$ is too long, resulting in incomplete overlap with the computation. This not only delays the unloading of $s_1$ and increases memory consumption, but also delays the backward pass of $s_0$, ultimately lengthening the end-to-end latency. To address Challenge 1, we introduce an \textbf{adaptive offloading} approach that effectively overlaps the offloading of activations with computation (Section \ref{sec:adaptive_offloading}). It refines the offloading ratio by \textit{sequence-aware offloading} and optimizes the selection of offloaded activation by a \textit{two-level activation management} strategy, as depicted in the middle part of Figure \ref{fig_system_overview}. Each subsequence tensor is divided into multiple tensor units, categorized based on their runtime lifespan and designated for either offloading or retention. At the intra-unit level, we implement two-level activation management to handle activations of varying lifespans. At the inter-unit level, we apply sequence-aware offloading, selectively deciding whether to offload tensors to CPU or retain them in GPU memory. Adaptive offloading optimizes training efficiency and reduces peak memory consumption from 7 to 5 units.

      \item \textsl{\bfseries Adaptive Pipeline}. Despite improvements, pipeline bubbles remain an issue, as shown in Challenge 2 in the middle part of Figure \ref{fig_system_overview}. To further enhance pipeline efficiency, we introduce an \textbf{adaptive pipeline} mechanism.  It applies a \textit{heuristic solver} to determine the ideal number of subsequences $N$. To reduce pipeline bubbles, this adaptive pipeline incorporates \textit{multiplexing sequence partitioning}, which refines subsequence partitioning to minimize resource waste without increasing memory overhead, as illustrated in the bottom part of Figure \ref{fig_system_overview}. By leveraging traditional sequence parallelism \cite{DeepSpeedUlysses}, we further partition the subsequence around pipeline bubbles into smaller sub-subsequences, distributing them across GPUs for parallel computation. For instance, $s_0$ in stage 0 is split into two parts, allowing forward and backward passes to execute in parallel at the beginning and end of runtime, thereby reducing bubble time in stage 1. Similarly, $s_3$ in stage 1 is partitioned and computed in parallel, reducing bubble time in stage 0. \sysname{} automatically recognizes those subsequences near the bubble and re-segment them without incurring additional memory overhead.

\end{itemize}

\noindent \textbf{Workflow}. Given a specific sequence length and available hardware resources, \sysname{} operates the adaptive offloading and adaptive pipeline as follows. (1) It applies the FLOPs-based partitioning policy for balanced computation across pipeline stages instead of the length-based partition policy. Specifically, the adaptive pipeline adopts the \textit{heuristic solver} to determine the optimal values for $N$ and $PP$. (2) The adaptive offloading approach then leverages \textit{sequence-aware offloading} to compute the offloading ratio for each subsequence and manages corresponding activations via \textit{two-level activation management}. (3) If the pipeline scheduling still exhibits excessive bubble ratios, the adaptive pipeline applies \textit{multiplexing sequence partitioning} to utilize these bubbles effectively, thereby enhancing training efficiency.

\section{Adaptive Offloading}
\label{sec:adaptive_offloading}
Traditional frameworks typically employ the same offloading strategy across layers or micro-batches. However, there exists an inherent imbalanced memory usage among different subsequences, which leads to imbalanced transmission time across GPUs and CPU that introduces extra transmission overhead. Besides, inside subsequences, different tensors may have different lifespans. \sysname{} allows each subsequence to find its best offloading strategy according to its following subsequence computation.

  To search for the best offloading strategy for different subsequences, we first analyze the inner structure of the transformer layer and split the layer activation into finer-grained activation units, each of which is a minimal group of tensors to be offloaded or saved in GPUs. Then, we construct the cost model and design the algorithms to find the best offloading ratio across different subsequences.

Below, we first introduce a \textbf{two-level activation management} scheme for inner tensors within a subsequence, and then present a \textbf{sequence-aware offloading} strategy. 

\subsection{Two-Level Activation Management}
\label{subsec_hierarchical_act_manager}
\begin{figure}[t!]
    \centering
    \includegraphics[width=\linewidth]{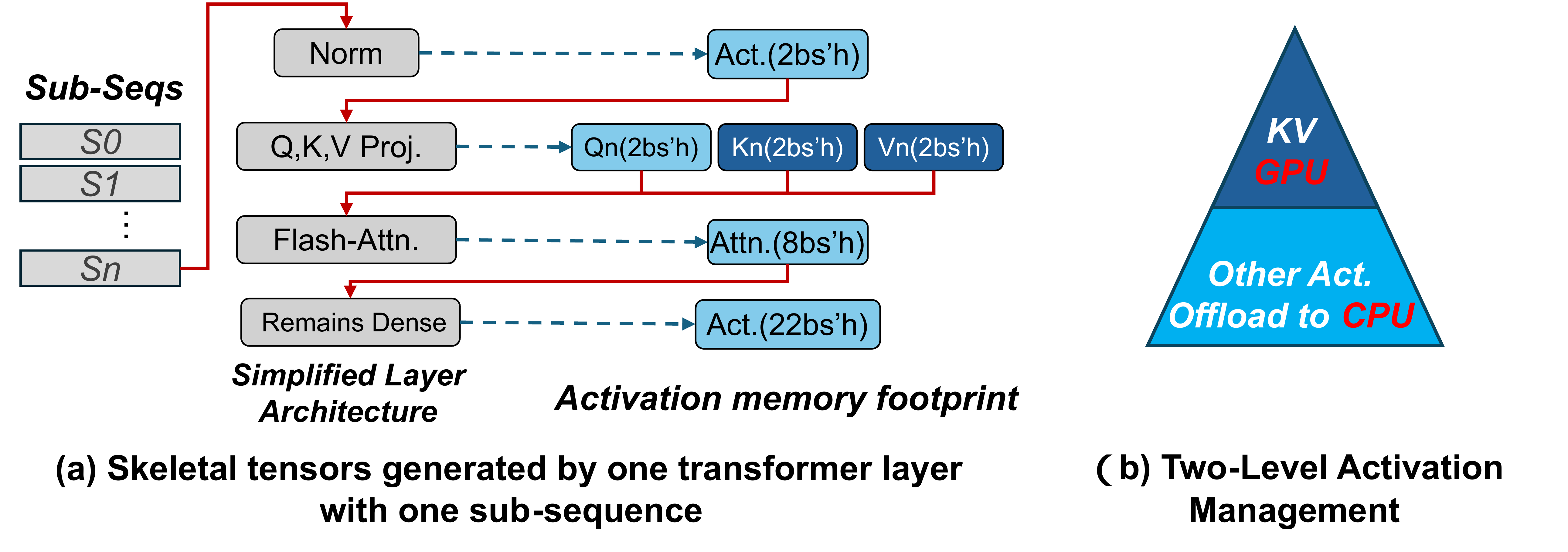}
    \caption{An example of computation of $s_N$ in a Transformer-based model and the skeletal tensors with its sizes.} 
    \label{fig_hierarchical_memory_manager}
\end{figure}

Figure \ref{fig_hierarchical_memory_manager} (a) gives an example of how to perform the computation of subsequence $s_N$ in a Transformer-based model, showing all skeletal tensors generated within a layer's forward pass of $s_N$ and their sizes. Note that, for generative LLMs, due to the casual mask, after attention computation, $Q_i$ will not be used in the forward pass anymore, while $K_i$ and $V_i$ need to participate in the following $Q_N$ computation, where $i<N$.  $K_0$ and $V_0$ are Type-0 tensors that are accessed continuously (Figure \ref{fig_unnecessary_offload}). Therefore, we cache $K_i$ and $V_i$ to GPU memory instead of offloading them to CPU. Caching all ($K$-$V$) pairs only consumes $\sum_{i=0}^{n}2Bs_{i}H=2BSH$ bytes of GPU memory, while we will partially offload the remaining $36BSH$ bytes of tensors to CPU. The offloading ratio is determined by our sequence-aware offloading strategy in Section \ref{subsec:sequence_aware_offloading}. The most important characteristic of the remaining activations is that they are needed by backward computation, so they just need to reside in GPU memory at least before the backward propagation of the subsequence begins. Figure \ref{fig_hierarchical_memory_manager} (b) shows the architecture of our activation management.

\subsection{Sequence-aware Offloading}
\label{subsec:sequence_aware_offloading}

This scheduling enables overlapping data transfer operations for the $(i-1)^{th}$ subsequence with the computation of the $i^{th}$ subsequence, thereby reducing end-to-end training time. However, incomplete offloading of the $(i-1)^{th}$ subsequence before the completion of the $i^{th}$ subsequence computation creates dual memory demands: the GPU must simultaneously store activation $A_i$ and maintain an offloading buffer for $A_{i-1}$. As demonstrated in Figure \ref{fig_system_overview} (top), this memory pressure peaks at 7 activation units when $S_0$ offloading persists through $S_2$ computation.

To address the imbalanced memory utilization, we introduce an adaptive offloading ratio $\alpha \in [0,1]$ that controls the proportion of activations transferred to host memory. Given a FLOPs-optimized subsequence partition $s=\{s_0,s_1,...,s_N\}$ with $s_0 \leq s_1 \leq ... \leq s_N$, we assign corresponding offloading ratios $\alpha=\{\alpha_0,\alpha_1,...,\alpha_N\}$ where $\alpha_0 \geq \alpha_1 \geq ... \geq \alpha_N$. The memory dynamics during pipeline stage $i$ follows:
\[
M_i = M_{i-1} + A_i - (\alpha_{i-1} \cdot A_{i-1})
\]
where $M_{i}$ is the current GPU memory consumption, $A_i$ is the activation volume of subsequence $i$, and $\alpha_{i-1} \cdot A_{i-1}$ is the offloaded portion of previous activation

We optimize $\alpha$ to maintain $\alpha_i \cdot A_i = M_{\text{threshold}}$, ensuring the offloading time $M_{\text{threshold}}/BW_{\text{D2H}}$ matches the balanced computation time $T_{\text{balanced\_comp}}$. This constraint yields the relationship $\alpha_{i-1} \cdot A_{i-1} = \alpha_i \cdot A_i$. Peak memory $M_{\text{max}}$ occurs when $\alpha_k = 1$ for final subsequence $k$, resulting in:
\[
M_{\text{max}} = M_{k-1} = (1 - \alpha_{k-1}) \cdot A_{k-1}
\]
This formulation enables systematic memory optimization through our proposed scheduling algorithm. 

\section{Adaptive Pipeline}
\label{sec:adaptive-pipeline}
\begin{table}[t]

    \centering
    
    \resizebox{\linewidth}{!}{
    \begin{tabular}{
        l
        >{\centering\arraybackslash}p{2.5cm} >{\centering\arraybackslash}p{2.5cm} 
        >{\centering\arraybackslash}p{2.5cm} >{\centering\arraybackslash}p{2.5cm}
    }
        \toprule
        \textbf{Component} & \textbf{Stage 0} & \textbf{Stage 1} & \textbf{Stage 2} & \textbf{Stage 3} \\
        \midrule
        Left Seq. IDs & $\{0,1,2\}$ & $\{0,1\}$ & $\{0\}$ & $\emptyset$ \\
        Steady Seq. IDs & $\{3,4,5,6,7\}$ & $\{2,3,4,5,6\}$ & $\{1,2,3,4,5\}$ & $\{0,1,2,3,4\}$ \\
        Right Seq. IDs & $\emptyset$ & $\{7\}$ & $\{6,7\}$ & $\{5,6,7\}$ \\
        \midrule
        Left SP Range & $\{0,1,2,3\}$ & $\{1,2,3\}$ & $\{2,3\}$ & $\emptyset$ \\
        Right SP Range & $\emptyset$ & $\{0,1\}$ & $\{0,1,2\}$ & $\{0,1,2,3\}$ \\
        \bottomrule
    \end{tabular}}
    \caption{Illustration of multiplexed sequence partitioning for $PP=4$ stages and $N=8$ subsequences. SP ranges denote GPU allocations for parallel computation.}
\label{tab:example_of_mux}
\end{table}
\subsection{Heuristic Solver}
Given the model, sequence length $s$, and the number of nodes, the objective is to search for a set of hybrid parallelism parameters $(SP, PP, N)$ that minimize the time per iteration $T$. To reduce the search space for performance tuning, we leverage the following heuristics:

\begin{itemize}[leftmargin=*,topsep=1pt, itemsep=2pt]
    \item \textbf{Avoid cross-node sequence parallelism}: Tensor parallelism incurs significantly larger communication overhead compared to pipeline parallelism within each device, making cross-node sequence parallelism inefficient.
    \item \textbf{Avoid cross-node pipeline parallelism}: The \texttt{P2P} communication bandwidth across nodes is typically limited, leading to performance degradation.
    \item \textbf{Maintain near-balanced workloads}: Empirical profiling (as shown in Figure \ref{fig_trade_off_chunksize}) suggests that the optimal workload size per layer ranges from 2K to 16K.
\end{itemize}

Following these guidelines, we restrict $N$ within the range of 2K to 16K per layer per device, facilitating efficient workload distribution and minimizing communication overhead.

\subsection{Multiplexing Sequence Partition}
To address significant pipeline bubbles caused by reduced values of $N$, we propose a latency optimization strategy termed \emph{multiplexing sequence partitioning}. Building on \sysname's adaptive offloading, our method introduces finer-grained partitioning of bubble-adjacent subsequences across distributed GPUs using the traditional Megatron sequence parallelism (Section \ref{subsec:distributed_parallel}). This approach effectively minimizes idle time in pipeline stages while avoiding additional memory overhead.

Formally, for a $PP$-stage pipeline processing $N$ subsequences, the forward pass of each stage consists of three computation phases:
\begin{itemize}[leftmargin=*,topsep=1pt, itemsep=2pt]
    \item \textbf{Left-SP}: Initial parallel computation phase leveraging distributed subsequences.
    \item \textbf{Steady}: Central computation phase optimized through adaptive offloading.
    \item \textbf{Right-SP}: Final parallel computation phase, continuing sequence parallelism.
\end{itemize}
Each phase is characterized by three properties: (1) execution paradigm, (2) subsequence identifier mapping, and (3) communication scope. The backward pass follows a similar yet asymmetric structure. While the Left-SP and Right-SP phases extend \sysname{} with secondary sequence partitioning, the Steady phase maintains the core adaptive offloading optimization.

For pipeline stage $i \in \{0, 1, \dots, PP-1\}$, its phase characteristics are defined as follows:

\begin{definition}[Subsequence Identification Mapping]
Given partition size $PP \in \mathbb{N}$, number of subsequences $N \in \mathbb{N}$, and stage index $i$, the phase-specific subsequence IDs $\mathcal{I}(i)$ are:
\begin{equation}
\mathcal{I}(i) =
\begin{cases}
\{x \in \mathbb{N}_0 \mid 0 \leq x \leq PP-1-i\} & \text{(Left-SP)} \\
\{x \in \mathbb{N}_0 \mid PP-1-i \leq x \leq N-i\} & \text{(Steady)} \\
\{x \in \mathbb{N}_0 \mid N-i \leq x \leq N-1\} & \text{(Right-SP)}
\end{cases}
\end{equation}
where $\mathbb{N}_0$ denotes non-negative integers.
\end{definition}

\begin{definition}[Inter-Stage Communication Scope]
The communication range $\mathcal{C}(i)$ for stage $i$ is:
\begin{equation}
\mathcal{C}(i) =
\begin{cases}
\{x \in \mathbb{N}_0 \mid i \leq x \leq PP-1\} & \text{(Left-SP)} \\
\{x \in \mathbb{N}_0 \mid 0 \leq x \leq PP-1\} & \text{(Steady)} \\
\{x \in \mathbb{N}_0 \mid 0 \leq x \leq i\} & \text{(Right-SP)}
\end{cases}
\end{equation}
\end{definition}

Figure \ref{fig_system_overview} illustrates this mechanism: Subsequence $s_0$ in Stage 0 undergoes partitioning, enabling parallel execution of forward and backward passes during the initial and final phases. Simultaneously, $s_3$ in Stage 1 is partitioned similarly, facilitating distributed computation and reducing bubble durations in adjacent stages. Our approach dynamically identifies critical subsequences near pipeline bubbles and performs memory-efficient repartitioning without reallocation.

Table \ref{tab:example_of_mux} illustrates this partitioning scheme for $PP=4$ pipeline stages and $N=8$ subsequences. The Left-SP phase processes edge subsequences near pipeline boundaries, while the Right-SP phase manages terminal computations. The Steady phase handles central subsequences using standard pipeline parallelism. Sequence parallelism ranges exhibit complementary patterns across stages, ensuring efficient resource utilization and minimizing cross-stage dependencies.

\section{Evaluation}
\begin{figure*}[]
    \centering
    \includegraphics[width=\linewidth]{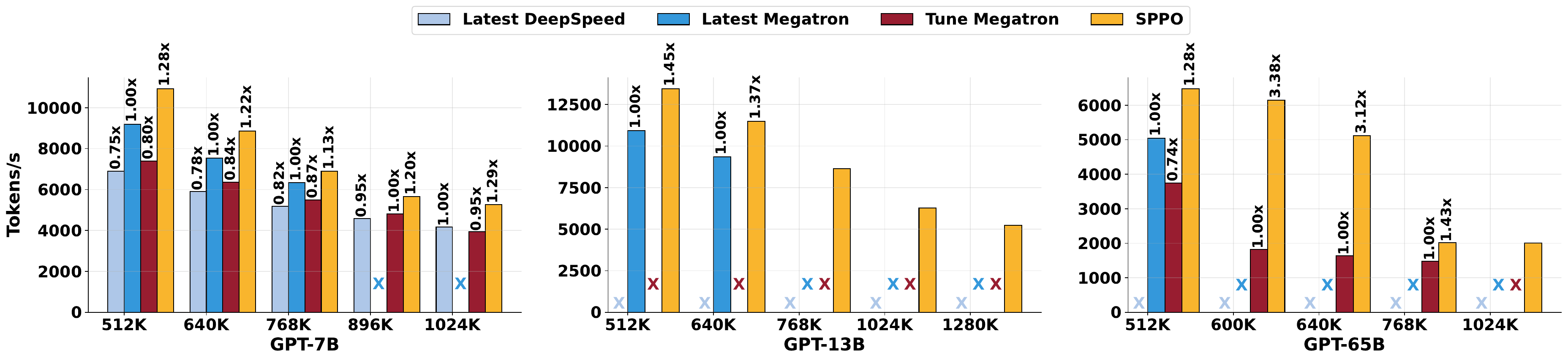}
    \caption{End-to-end evaluation results of training models of different sizes and sequence lengths.} 
    \label{eval:e2e}
\end{figure*} 

\noindent\textbf{Implementation.} \sysname{} is implemented using Python and CUDA and encompasses around 4000 lines of code (LOC) based on Megatron-LM. The codebase consists of 2845 LOC for \sysname{} optimized by adaptive offloading and 1155 LOC for multiplexing sequence partitioning. To achieve the highest bandwidth between GPU and host, we bind the Non-Uniform Memory Access node for each process and use page-locked memory for CPU buffers.

\noindent \textbf{Testbed.} \sysname{} undergoes a thorough evaluation in the training of Transformer-based models with the GPT architecture. The size of the models range from 7 billion to 65 billion, and detailed configurations can be found in Table~\ref{tab:min_num_gpus}. The training process occurs on a physical cluster comprising 16 GPU servers. Each server is equipped with 8 GPUs and 128 CPU cores, a total of 128 NVIDIA Ampere GPUs. Each GPU boasts 80GB of memory. The GPUs are interconnected through NVLink and NVSwitch, while inter-node communication is facilitated by four NVIDIA Mellanox 200 Gbps HDR InfiniBand. Each node has 2 TB of CPU memory, and the GPU-CPU communication bandwidth is 32 GB/s.

\noindent\textbf{Baselines and Metrics.} We benchmark \sysname{} against two state-of-the-art systems for long-sequence LLM training, namely DeepSpeed Ulysses \cite{DeepSpeedUlysses} and Megatron-LM \cite{nvidia3}. The evaluation metric is tokens per GPU per second (TGS) during training.  By default, we enable activation checkpointing for baselines, allowing them to support longer sequence lengths. We strengthen DeepSpeed Ulysses \cite{ZeRO} with ZeRO to reduce the memory footprint of weights, gradient, and optimizer states and with FPDT \cite{FPDT} to offload activations.

\begin{table}[h]
    \centering
    
    % \small % 字体比 small 更小
    \renewcommand{\arraystretch}{0.9} % 进一步减少行高
    \setlength{\tabcolsep}{1pt} % 减小列间距
    \resizebox{\linewidth}{!}{
    \begin{tabular}{
        c c c 
        >{\centering\arraybackslash}p{1.2cm} >{\centering\arraybackslash}p{1.2cm} 
        >{\centering\arraybackslash}p{1.2cm} 
        >{\centering\arraybackslash}p{1.2cm} >{\centering\arraybackslash}p{1.2cm} 
        c
    }
        \toprule
        \textbf{Model} & \textbf{\#GPUs} & \textbf{S} & \multicolumn{2}{c}{\makecell{Mega. \\ \& SPPO}} & \multicolumn{1}{c}{\makecell{SPPO}} & \multicolumn{2}{c}{\makecell{Tune \\Mega.}} & \textbf{DS} \\
        \cmidrule(lr){4-5} \cmidrule(lr){6-6} \cmidrule(lr){7-8}
        & & & \textbf{SP} & \textbf{PP} & \textbf{N} & \textbf{SP} & \textbf{PP} & \textbf{SP} \\
        \midrule
        GPT-7B & 32 & 512K & 8 & 4 & 32 & 32 & 1 & 32 \\
         &  & 640K &  &  & 64 &  &  &  \\
         &  & 768K &  &  & 80 &  &  &  \\
         &  & 896K &  &  & 128 &  &  &  \\
         &  & 1024K &  &  & 160 &  &  &  \\
        \midrule
        GPT-13B & 64 & 512K & 8 & 8 & 32 & 8 & 8 & - \\
         &  & 640K &  &  & 48 &  &  &  \\
         &  & 768K &  &  & 64 &  &  &  \\
         &  & 1024K &  &  & 80 &  &  &  \\
         &  & 1280K &  &  & 160 &  &  &  \\
        \midrule
        GPT-65B & 128 & 512K & 16 & 8 & 32 & 64 & 2 & - \\
         &  & 600K &  &  & 64 &  &  &  \\
         &  & 640K &  &  & 80 &  &  &  \\
         &  & 768K &  &  & 96 &  &  &  \\
         &  & 1024K &  &  & 128 &  &  &  \\
        \bottomrule
    \end{tabular}
    }
    \caption{Configurations for various GPT models.}
    \label{tab:config_in_eval}
\end{table}

\begin{figure*}[]
    \centering
    \includegraphics[width=\linewidth]{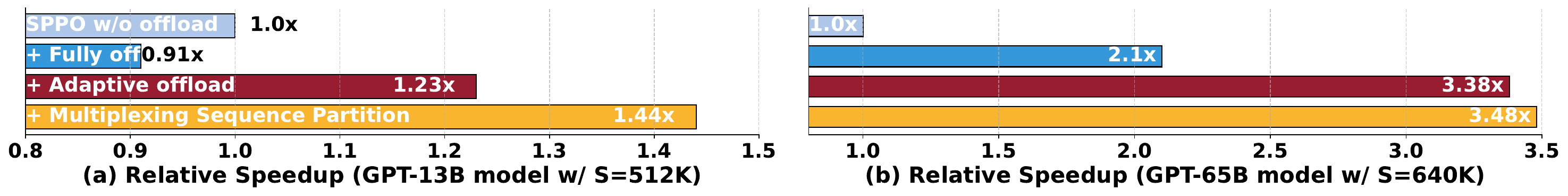}
    \caption{Breakdown analysis} 
    \label{eval:breakdown}
\end{figure*}

\subsection{End-to-End Evaluation}
Figure \ref{eval:e2e} illustrates the training throughput across various LLMs and sequence lengths. For fairness, we optimize the parallelism strategy for each system, as reported in Table \ref{tab:config_in_eval}.

\noindent\textbf{Throughput and Scalability:} In our physical testbed, \sysname{} consistently outperforms DeepSpeed-Ulysses and Megatron-LM across various sequence lengths and model sizes, achieving speedups of ranging from 1.13$\times$ to 3.38$\times$. A key advantage of \sysname{} is its ability to handle ultra-long sequences without out-of-memory (OOM) issues. In contrast, Megatron-LM struggles with sequences exceeding 896K for GPT-7B, while DeepSpeed-Ulysses fails to support sequence lengths beyond 512K for GPT-13B and GPT-65B.

% \sysname{} scales seamlessly of up to 1024K and beyond.

% This is particularly evident for longer sequences, where \sysname{} achieves significant speedups, ranging from 1.13$\times$ to 3.38$\times$ over the baselines. The ability of \sysname{} to handle longer sequences without encountering out-of-memory (OOM) issues is a major advantage. For example, while Megatron-LM struggles with sequences beyond 896K for GPT-7B, \sysname{} scales seamlessly of up to 1024K and beyond.

\noindent\textbf{Memory Efficiency:} \sysname{} mitigates the memory pressure through nearly zero-overhead offloading and pipeline parallelism. It avoids costly activation recomputation, a limitation that significantly impacts Megatron-LM under the same parallelism strategy. For GPT-65B, \sysname{} supports sequence lengths of up to 1024K, whereas DeepSpeed-Ulysses is limited to 512K and Megatron-LM to 768K. This demonstrates \sysname{}'s superior memory management capabilities.

\noindent\textbf{Model-Specific Performance:} For GPT-7B, under the same parallelism strategy as the latest Megatron-LM, \sysname{} achieves a 1.13$\times$ to 1.29$\times$ speedup. At sequence lengths beyond 896K, Megatron-LM encounters OOM issues, while \sysname{} continues to deliver superior throughput. For GPT-13B, \sysname{} supports sequence lengths of up to 1280K, whereas Megatron-LM is limited to 768K. DeepSpeed-Ulysses cannot train GPT-13B at all due to model architectural constraints. For GPT-65B, \sysname{} achieves remarkable speedups of 3.38$\times$ and 3.12$\times$ over Megatron-LM at sequence lengths of 600K and 640K, respectively. DeepSpeed-Ulysses, on the other hand, is unable to scale effectively for GPT-65B beyond 512K.

\noindent\textbf{Limitations: }While \sysname{} shows superior performance, it is important to note that the baseline systems (DeepSpeed-Ulysses and Megatron-LM) have their own strengths in specific scenarios. For instance, Megatron-LM performs well for shorter sequence lengths (e.g., a GPT-7B model with a sequence length of 768K) despite employing activation recomputation. In such cases, \sysname{} offers only marginal improvements, as the computational workload of the GPT-7B model is insufficient to fully exploit offloading overlap.

% The evaluation does not account for potential trade-offs in hardware utilization or energy efficiency, which could be explored in future work.

\subsection{Speedup Breakdown}
To gain deeper insights into the key optimizations contributing to \sysname{}'s benefits, we conduct a performance breakdown analysis to present the impact of each key technique. Figure \ref{eval:breakdown} presents the normalized speedup performance against Megatron-LM. We have three key observations. 

% conduct an incremental performance evaluation by progressively enabling the techniques proposed in \sysname{}. The results, normalized against Megatron-LM, are illustrated in Figure \ref{eval:breakdown}. Three key observations emerge from the analysis:

First, full CPU offloading alone does not always improve performance. While CPU offloading can optimize GPU memory efficiency, activation fully offloading overhead is non-negligible in certain scenarios, as discussed in Section \ref{sec:adaptive_offloading} and shown in Figure \ref{eval:breakdown}(a). \sysname{} only attains a relative speedup of 0.91 compared to SPPO w/o offload in a GPT-13B model with a sequence length of 512K. However, full CPU offloading benefits the parallelism strategies that exhibit high communication efficiency but poor memory efficiency, as demonstrated in Figure \ref{eval:breakdown}(b). It achieves a relative speedup of 2.1 $\times$ in the GPT-65B model with a sequence length of 640K.

% While full offloading can improve GPU memory efficiency, it introduces additional latency in end-to-end execution, as discussed in Section \ref{sec:adaptive_offloading} and shown in Figure \ref{eval:breakdown}(a). For instance, in the GPT-13B model with a sequence length of 512K, full offloading results in a relative speedup of only 0.91x compared to the baseline (SPPO w/o offload). However, for parallelism strategies that exhibit high communication efficiency but poor memory efficiency, selective offloading can enhance performance, as demonstrated in Figure \ref{eval:breakdown}(b). For example, in the GPT-65B model with a sequence length of 640K, full offloading achieves a relative speedup of 2.1x, indicating its effectiveness in specific scenarios.

Second, adaptive offloading consistently improves training and memory efficiency over different scenarios. Empirically, it brings a relative speedup of 1.23 $\times$ and 3.38 $\times$ for GPT-13B and GPT-65B, respectively, indicating its superiority in handling varying model sizes and sequence lengths.

% Across all evaluated workloads, adaptive offloading delivers a noticeable speedup, demonstrating its effectiveness in dynamically optimizing memory usage. For the GPT-13B model, adaptive offloading achieves a relative speedup of 1.23x, while for the GPT-65B model, it achieves an even more significant speedup of 3.38x. This highlights the robustness of adaptive offloading in handling varying model sizes and sequence lengths.

Third, multiplexing sequence partitioning significantly enhances training efficiency. Its benefits are particularly pronounced when the bubble ratio is high, which is determined by the number of subsequences ($N$) and pipeline stages ($P$) used during training. MSP achieves a relative speedup of 1.44 $\times$ for the GPT-13B model and a remarkable speedup of 3.48 $\times$ for the GPT-65B model. 

Overall, each optimization technique plays a pivotal role in \sysname{}, collectively contributing to the significant speedup demonstrated by our empirical results.

% In summary, each of the optimizations incorporated into \sysname{} plays a crucial role in enhancing system performance. Offloading, adaptive offloading, and multiplexed sequence partitioning collectively contribute to substantial speedup, as evidenced by the experimental results. While full offloading may not always be beneficial, adaptive offloading and MSP consistently deliver significant performance improvements across different model sizes and sequence lengths. These findings highlight the importance of a balanced and context-aware approach to optimizing large-scale model training.

\subsection{Sequence Length Scalability}
\begin{figure}[]
    \centering
    \includegraphics[width=\linewidth]{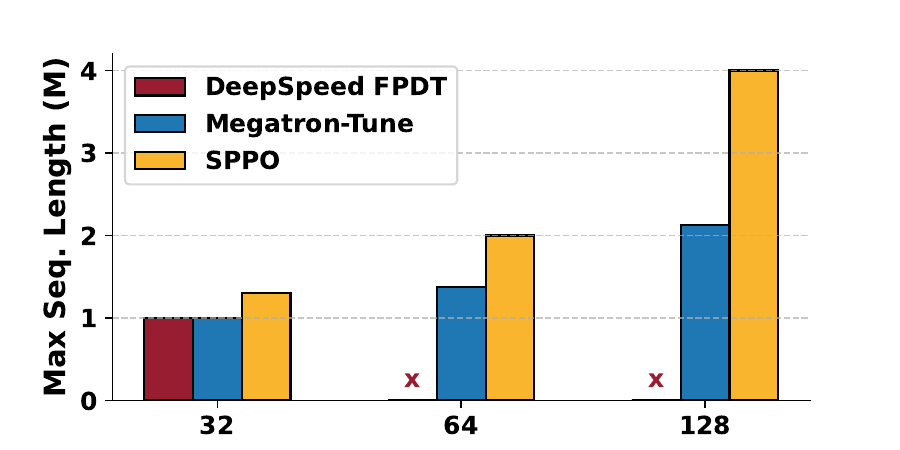}
    \caption{Scaling maximum sequence length with different number of GPUs in training GPT-7B model.} 
    \label{eval:maxlen}
\end{figure}

To analyze the sequence length scalability of \sysname{} with fixed GPU resources, we progressively increase the per-GPU sequence length from 1K to 10K tokens until experiencing OOM issues. We adopt activation checkpointing in each layer for DeepSpeed and Megatron-LM. As shown in Figure~\ref{eval:maxlen}, \sysname{} demonstrates clear advantages in sequence length scalability compared to the other two baselines.

Specifically, DeepSpeed-Ulysses partitions computation along with attention heads, making it less effective for models with a limited number of heads. For instance, the restricted number of heads in GPT-7B becomes a scalability bottleneck for DeepSpeed-Ulysses. This explains why DeepSpeed-Ulysses gets a sharp decline in the maximum supported sequence length, dropping from 1K at 32 GPUs to 0 at 64 GPUs. Megatron-Tuned exhibits partial scalability (1$\times$, 1.375$\times$, 2.13$\times$ at 32/64/128 GPUs). The activation recomputation hinders the linear scaling in sequence length. 

In contrast, \sysname{} is not constrained by the number of attention heads, achieving near-linear scalability. It can support sequence lengths of $1.3\times$, $2\times$, and 4$\times$ the baseline at 32, 64, and 128 GPUs, respectively. At 128 GPUs, \sysname{} outperforms Megatron-Tuned by 88\% (4$\times$ vs. 2.13$\times$). Moreover, \sysname{} achieves higher speedup at more GPUs (4$\times$ at 128 GPUs vs. 2$\times$ at 64 GPUs), indicating its ability to reduce communication overhead and enhance memory efficiency.

In sum, our empirical analysis of sequence length scalability highlights the critical role of parallelism strategies in long-sequence training. By decoupling from head-based partitioning, \sysname{} can flexibly adjust the configurations for parallelism strategies, enabling the training of sequences exceeding multi-million tokens. In contrast, existing LLM training systems face significant challenges in handling such extreme-scale scenarios.

\section{Related Work}
\noindent\textbf{Systems for long-sequence training.} To address memory and computational limits, ColossalAI-SP \cite{ColosslaiSP} introduces sequence segmentation and parallelism alongside tensor and pipeline parallelism. Ring Attention \cite{liu2023blockwise,BPT2} further improves efficiency by using blockwise self-attention to distribute long sequences across devices while overlapping key-value communication. LightSeq \cite{li2023lightseq} optimizes long-sequence modeling via load balancing and re-materialization-aware checkpointing. Some approaches integrate efficient self-attention mechanisms like FlashAttention \cite{flashatten1,flashatten2}. Megatron-LM \cite{nvidia3} applies sequence parallelism selectively in Dropout and LayerNorm to reduce activation redundancy, while DeepSpeed-Ulysses \cite{DeepSpeedUlysses} employs all-to-all collective communication to avoid increasing overhead with sequence length. Hybrid Sequence Parallelism \cite{loongtrain} combines Ring Attention and DeepSpeed-Ulysses to enhance scalability and efficiency.

\noindent\textbf{Activation recomputation and swapping.} Capuchin \cite{capuchin} reduces memory footprint by combining recomputation and swapping, considering tensor access patterns. MegTaichi \cite{megtaichi} co-optimizes tensor partitioning, while Coop \cite{coop} minimizes memory fragmentation in recomputation. These approaches do not fully leverage LLM training characteristics for optimal overlapping and fragmentation reduction. ZeRO-Offload \cite{zeroOffload} offloads optimizer states to host memory, and vDNN \cite{vdnn} schedules prefetching and offloading for better overlap. SuperNeurons \cite{superneurons} balances offloading and recomputation by offloading compute-heavy activations while recomputing lighter ones. ZeRO-Infinity \cite{ZeRO-Infinity} utilizes NVMe SSDs for large-scale training but suffers from high CPU-GPU communication costs. With modern GPUs, effectively overlapping computation with communication remains a challenge.

\noindent\textbf{Reducing pipeline bubbles.} Several efficient micro-batch scheduling algorithms have been proposed to mitigate the pipeline bubbles in deep learning training. GPipe \cite{GPipe} introduces a fill-drain schedule but suffers from pipeline inefficiencies due to warm-up and cool-down phases. PipeDream \cite{PipeDream} employs a 1F1B schedule to reduce bubbles by executing the backward pass immediately after the forward pass of a micro-batch. DAPPLE \cite{dapple} improves upon this with an early backward schedule, while Interleaved 1F1B \cite{Megatron-LM1} extends 1F1B with multi-stage assignments per GPU. Chimera \cite{Chimera,Hanayo} implements a bidirectional pipeline with weight duplication to further reduce bubbles. Zero Bubble \cite{zerobuble} mitigates bubbles by splitting backward computation, leveraging 1F1B scheduling and parameter gradient computation. Breadth-First \cite{breadthfirst} processes all micro-batches simultaneously in looping pipeline placement to minimize communication overhead. TeraPipe \cite{TeraPipe} and Seq1F1B \cite{seq1f1b} focus on sequence-level partitioning to balance memory and efficiency. DynaPipe \cite{dynapipe,tessel} introduces adaptive scheduling for multi-task LLM training, optimizing memory usage and communication planning. DISTMM \cite{distmm} launches doubled micro-batches to circumvent dependency barriers in multi-modal training, while GraphPipe \cite{graphpipe} preserves DNN graph topology for concurrent execution, improving pipeline efficiency and memory consumption.
\section{Conclusion}
In this work, we introduce Adaptive Sequence Pipeline Parallel Offloading (\sysname{}), a novel framework designed to enhance the efficiency of long-sequence LLM training by addressing memory and computational resource limitations. By leveraging adaptive offloading and optimized pipeline scheduling, \sysname{} effectively balances memory usage and training speed, overcoming the inefficiencies of existing methods. Our experimental results demonstrate significant performance improvements, achieving up to 3.38× higher throughput compared to state-of-the-art frameworks while reducing GPU resource requirements. These advancements pave the way for scalable and efficient training of LLMs with extremely long input sequences, enabling broader applications across AI research and industry.

%-------------------------------------------------------------------------------
% \bibliographystyle{unsrt}
\bibliographystyle{ACM-Reference-Format}
\bibliography{reference}

\end{document}